\documentstyle[epsfig]{mn}

\newif\ifAMStwofonts

\ifoldfss
  \ifCUPmtlplainloaded \else
    \NewTextAlphabet{textbfit} {cmbxti10} {}
    \NewTextAlphabet{textbfss} {cmssbx10} {}
    \NewMathAlphabet{mathbfit} {cmbxti10} {} 
    \NewMathAlphabet{mathbfss} {cmssbx10} {} 
  \fi
  \ifAMStwofonts
    \ifCUPmtlplainloaded \else
      \NewSymbolFont{upmath} {eurm10}
      \NewSymbolFont{AMSa} {msam10}
      \NewMathSymbol{\upi}     {0}{upmath}{19}
      \NewMathSymbol{\umu}     {0}{upmath}{16}
      \NewMathSymbol{\upartial}{0}{upmath}{40}
      \NewMathSymbol{\leqslant}{3}{AMSa}{36}
      \NewMathSymbol{\geqslant}{3}{AMSa}{3E}

    \fi
  \fi
\fi 

\ifnfssone
  \newmathalphabet{\mathit}
  \addtoversion{normal}{\mathit}{cmr}{m}{it}
  \addtoversion{bold}{\mathit}{cmr}{bx}{it}
  \newmathalphabet{\mathbfit} 
  \addtoversion{normal}{\mathbfit}{cmr}{bx}{it}
  \addtoversion{bold}{\mathbfit}{cmr}{bx}{it}
  \newmathalphabet{\mathbfss} 
  \addtoversion{normal}{\mathbfss}{cmss}{bx}{n}
  \addtoversion{bold}{\mathbfss}{cmss}{bx}{n}
  \ifAMStwofonts
    \ifCUPmtlplainloaded \else
      \UseAMStwoboldmath
      \makeatletter
      \new@mathgroup\upmath@group
      \define@mathgroup\mv@normal\upmath@group{eur}{m}{n}
      \define@mathgroup\mv@bold\upmath@group{eur}{b}{n}
      \edef\UPM{\hexnumber\upmath@group}
      \new@mathgroup\amsa@group
      \define@mathgroup\mv@normal\amsa@group{msa}{m}{n}
      \define@mathgroup\mv@bold\amsa@group{msa}{m}{n}
      \edef\AMSa{\hexnumber\amsa@group}
      \makeatother
      \mathchardef\upi="0\UPM19
      \mathchardef\umu="0\UPM16
      \mathchardef\upartial="0\UPM40
      \mathchardef\leqslant="3\AMSa36
      \mathchardef\geqslant="3\AMSa3E
    \fi
  \fi
\fi 

\ifnfsstwo
  \DeclareMathAlphabet{\mathbfit}{OT1}{cmr}{bx}{it}
  \SetMathAlphabet\mathbfit{bold}{OT1}{cmr}{bx}{it}
  \DeclareMathAlphabet{\mathbfss}{OT1}{cmss}{bx}{n}
  \SetMathAlphabet\mathbfss{bold}{OT1}{cmss}{bx}{n}
  \ifAMStwofonts
    \ifCUPmtlplainloaded \else
      \DeclareSymbolFont{UPM}{U}{eur}{m}{n}
      \SetSymbolFont{UPM}{bold}{U}{eur}{b}{n}
      \DeclareSymbolFont{AMSa}{U}{msa}{m}{n}
      \DeclareMathSymbol{\upi}{0}{UPM}{"19}
      \DeclareMathSymbol{\umu}{0}{UPM}{"16}
      \DeclareMathSymbol{\upartial}{0}{UPM}{"40}
      \DeclareMathSymbol{\leqslant}{3}{AMSa}{"36}
      \DeclareMathSymbol{\geqslant}{3}{AMSa}{"3E}
    \fi
  \fi
\fi 

\ifCUPmtlplainloaded \else
  \ifAMStwofonts \else 
    \def\upi{\pi}
    \def\umu{\mu}
    \def\upartial{\partial}
  \fi
\fi

\title[CMB power spectrum with EM]{CMB power spectrum estimation and map
reconstruction with the
Expectation - Maximization algorithm}
\author[E. Mart\'\i nez-Gonz\'alez et al.]
   { E. Mart{\'\i}nez-Gonz{\'a}lez$^1$, J.M  Diego$^2$, P. Vielva $^{1,3}$ and
     J. Silk$^2$. \\
   $^1$Instituto de F{\'\i}sica de Cantabria, Consejo Superior de
     Investigaciones Cient{\'\i}ficas-Universidad de Cantabria, \\
     Avda. Los Castros s/n, 39005 Santander, Spain\\ 
   $^2$Astrophysics Dept. NAPL, Keble Road, Oxford OX1 3RH,  UK\\ 
   $^3$Departamento de F{\'\i}sica Moderna. Universidad de Cantabria, 
     Avda. Los Castros s/n, 39005 Santander, Spain\\}

\pagerange{\pageref{firstpage}--\pageref{lastpage}}

\begin{document}

\maketitle

\label{firstpage}
\begin{abstract}
We apply the iterative Expectation-Maximization algorithm (EM) to estimate the 
power spectrum of the CMB from multifrequency microwave maps. In addition, we 
are also able to provide a reconstruction of the CMB map. 
By assuming that the combined emission of the foregrounds plus the 
instrumental noise is Gaussian distributed in Fourier space, we have
simplified the EM procedure finding an analytical expression for the 
maximization step. By using the simplified expression the CPU time can be 
greatly reduced.
We test the stability of our power spectrum estimator with realistic 
simulations of 
Planck data, including point sources and allowing for spatial variation
of the frequency dependence of the Galactic emissions. 
Without prior information about any of the components, our new estimator 
can recover the CMB power spectrum up to scales $l\approx 1500$ with less than 10 \% 
error. This result is significantly improved if the brightest point sources are removed before 
applying our estimator. In this way,
the CMB power spectrum can be recovered up to
$l\approx 1700$ with 10 \% error and up to $l\approx 2100$ with 50 \% error.
This result is very close to the one that would be obtained in the ideal case
of only CMB plus white noise, for which all our assumptions are satisfied.
Moreover, the EM algorithm also provides an straightforward mechanism to reconstruct the
CMB map. The recovered cosmological signal shows a high degree of correlation (r = 0.98) with
the input map and low residuals.
\end{abstract}


\section{Introduction}\label{section_introduction}

Undergoing CMB experiments like BOOMERanG
(Netterfield et al. 2002, Rulh et
al. 2002), MAXIMA
(Hanany et al. 2000), DASI (Halverson et al. 2002), VSA
(Grainge et al. 2002), CBI (Mason et al. 2002), ACBAR 
(Kuo et al. 2002), Archeops (Benoit et al. 2002) and MAP
as well as future ones (Planck) will reveal 
with unprecedented quality the primordial matter fluctuations responsible 
for the CMB. 
Very recently the first detection of the E-mode polarization
in the CMB has been claimed (DASI, Kovac et al. 2002), which independently
supports the structure formation models via
gravitational instability.
Thanks to these experimental results the 
fundamental cosmological parameters can be determined with good accuracy.

These experiments will measure also the emission coming from our own Galaxy at 
mm frequencies as well as the emission due to other galaxies in the same wavebands. 
On the other hand, galaxy clusters will distort the CMB radiation with an 
intensity which is proportional to the total mass of the cluster times its temperature. 
All these components will generate a confusion limit which, together with the intrinsic 
detector noise, will make very difficult to disentangle which percentage of the 
observed emission is due to the CMB or to the other components. 
Several component separation methods have been proposed so far in the
literature: multifrequency
Wiener filter (MWF, Tegmark and Efstathiou 1996, Bouchet and Gispert 1999), 
maximum entropy methods (MEM, Hobson et al. 1998, Vielva et al. 2001b;
extended to the sphere in Stolyarov et
al. 2002, Barreiro et al. 2003), independent component analysis
(ICA, Baccigalupi et al. 2000, Maino et al. 2002), blind Bayes (Snoussi et al. 2001).  \\
Although there are some similarities and differences between the previous methods, all of them 
share a common aspect: they try to separate all the components
simultaneously. For this purpose,
these methods usually need to assume some {\it a priori} information about the components. 
Thus, if the different components have different frequency dependencies, then,  
by examining the data at different frequencies it is possible to distinguish 
(at a certain level) the different components. 
Or, if we know the correlation function of the components (or their power spectrum) and 
each one is significantly different from each other, then, it is also possible to 
distinguish (again at a certain level) the components since their modes will behave 
differently in the Fourier space. 
If one knows both, the frequency dependence and the power spectrum, then the 
component separation improves dramatically since now it is possible to combine the 
information in the different channels by correlating the Fourier modes with a 
correlation given essentially by the frequency dependence of each component 
and its power spectrum. 
This approach has proved to be very useful if the frequency dependence and power spectrum 
of the components are known. \\
It is interesting to explore what kind of information can be obtained when no 
prior information is assumed about none of the components. In some cases we simply do not 
know the priors (as it happens in the case of the free-free emission or the spinning 
dust). On the other hand, if we assume something 
about the components which is not accurate but an approach, then we are intrinsically 
introducing some bias in our final result. If, after the component separation, we end up 
with a biased estimate on, at least, one of the components, this will have an effect 
on some (if not all) of the other components which must {\it absorb} that bias 
in order to obey the constraint that the sum of all the components must be equal to 
the original data. This very last point is one of the risks of the simultaneous component 
separation methods. \\
Another assumption usually made is that the components 
are, one from each other, statistically independent. This assumption is not true for several 
of the components (like the Galactic ones). This assumption can be a source of 
systematic errors.
Other typical assumption is that the probability distribution function (pdf) 
of the individual components is a Gaussian. 
This is false for components like the point sources, the 
Sunyaev-Zeldovich (SZ) and the Galactic emissions where the true pdf 
has a bell-shape with a long tail towards positive values (negative
for the SZ for frequencies below 217 GHz). This assumtion will be somewhat 
relaxed in this work by assuming that the combined foreground emission plus 
instrumental noise is Gaussian distributed in Fourier space, instead of 
assuming it for each individual component. We will study the effect of this 
assumption at different scales. At those scales where the assumption is less 
justified solutions will be proposed to reduce the effect.

In a recent paper (Delabrouille et al. 2002) the authors have used
the Expectation-Maximization (EM)  algorithm to minimize a spectral matching 
criterion. They consider the problem of component separation with four 
components, CMB, SZ effect, 
dust and instrumental noise. In that work, the authors have shown that this 
is an interesting alternative to estimate jointly the frequency 
dependence and spatial power spectra of the components. This method allows to 
introduce certain degrees of freedom 
or {\it free parameters} in the problem. These free parameters can be 
determined after iterating an expectation-maximization process.

In the work presented here, we will explore this direction but in a much more
simplified manner focusing only on the estimation of the power spectrum and 
the map reconstruction of the CMB but including 
more components than in the previous work and doing no assumptions at all 
about the power spectrum or frequency dependence of any of the components. 
In addition to the diffuse foregrounds, we also will study the effect of 
point sources which were not considered in the previous work.

The approach presented in this paper can be extended to determine jointly the CMB and
the SZ components. The physics of these two emissions is well known, in
particular their frequency dependence. Therefore, no prior assumptions about
their frequency behaviour is required. In addition, these two components --together with
the point sources one-- are the most important signals from the cosmological point
of view. Several works have already been presented to determine the SZ 
(Herranz et al. 2001b,c; Diego et al. 2002) and the
point sources emissions (Tegmark \& Oliveira-Costa 1998; Cay{\'o}n et al. 2000; 
Sanz et al. 2001; Vielva et al. 2001, 2003; Chiang et al. 2002;
Hobson \& McLachlan 2002) from microwave images. Our aim for the future will be
to develop a method based on the EM algorithm combined with a point source detection
technique, to recover simultaneously the three cosmological emissions.

The paper is organized as follows. In Section 2, we summarize the key
ideas behind the EM algorithm. In Section 3 we apply the EM algorithm to the 
problem of determining an estimator of the CMB power
spectrum. The main results are shown in Section 4. These results are
compared with the ones which would be obtained in the ideal case when only CMB 
and white noise are considered. Finally, the conclusions are given in
Section 5.

\section{The Expectation-Maximization algorithm}\label{sect_EM}

In this section we will present an alternative method which will be useful to get a robust 
estimate of the power spectrum of the CMB. We will also present a single iterative expression 
for the new estimator of the CMB power spectrum. This estimator can be used directly with 
multifrequency microwave data to give a fast, accurate and robust estimate of the power spectrum of the CMB.\\
EM is an algorithm useful when there is a {\it many-to-one} 
mapping. That is, several variables are combined together 
to give one observed quantity. 
This is exactly our problem where the complete data are just the 
different components (Galactics, extragalactics, CMB and noise) and the observed quantity 
is just the {\it projection} of all of them in one of the channels (data on the receiver). 
EM allows to introduce degrees of freedom (or free parameters) in the pdf of the complete data 
and then look for the best parameters by maximising the expected value of that pdf 
given the observations. \\
The advantage of this algorithm in our problem is obvious. Since part of the information 
about the components is unknown, we can parametrise this unknown information as free parameters 
in the pdf of the complete data. 
Then, the maximization process will give us the best set of free parameters given the data. 

The EM algorithm (Dempster et al. 1977) provides an iterative procedure for 
computing maximum 
likelihood estimates (MLEs) in situations where the observed data
vector, $\bf{d}$, is viewed as being {\it incomplete}. 
$\bf{d}$ is an observable function of the
so-called {\it complete data} ${\bf x}$ (see e.g. McLachlan and Krishnan 1997).
Let's denote by  ${\bf f_i( d| p)}$ the pdf's of the incomplete data $\bf d$ 
and by ${\bf f(x|p)}$  the pdf of ${\bf x}$.
The vector ${\bf p=(p_1, .... ,p_N)}$ will contain the $N$ unknown parameters. 
The complete-data log likelihood function that could be formed for
${\bf p}$ if ${\bf x}$ were fully observable is given by 
\begin{equation}
\log L(\bf p)= \log f(\bf x| \bf p). 
\end{equation}
Instead of observing ${\bf x}$, we observe ${\bf d}$,
with many-to-one 
mapping from ${\bf x}$ to ${\bf d}$ ({\it i.e} ${\bf d = d(x)}$). 
It follows that 
\begin{equation}
{\bf{f_i( d| p) = \int_{X_d}} {\bf{f(x| p) dx}}}
\end{equation}
where ${\bf X_d}$  is the sub-space of ${\bf x}$ defined by
${\bf d = d(x)}$.\\ 
The EM algorithm approaches the problem of solving the
incomplete-data likelihood, ${\log \mathcal{L}\rm(\bf p)= \log f_i( d|
p)}$, indirectly by proceeding iteratively in terms of the
complete-data log likelihood function, $logL({\bf p})$. As it is
unobservable, it is replaced by its conditional expectation given
${\bf d}$, using the current fit for ${\bf p}$.
More specifically, the EM algorithm consists of two steps: Expectation (E-step)
and Maximization (M-step). On the j+1 iteration, these steps are as follows:\\

\noindent
$\bullet$ E-step: Calculate the quantity $Q({\bf p| \bf p^j})$, defined as 
$Q({\bf p| \bf p^j}) = E\{ \log L({\bf p ) | \bf d, \bf p^j}\} $.  \\

\noindent
$\bullet$ M-step: Choose ${\bf p^{j+1}}$ 
to be the value that maximises $Q({\bf p| \bf p^j})$; that is,
$Q({\bf p^{j+1}| \bf p^j}) = \max [Q({\bf p| \bf p^j})]$.\\

That is, in the first step we compute the expected value of the log of the complete 
pdf given the data and an estimate of the parameters (${\bf p^j}$). 
In the second step we look for the values of the parameters which maximise the 
previous expected value. At this step, we have to maximise with respect to each one 
of the parameters ($p^j(i)$). This can be done by just setting the first derivative 
of $Q({\bf p; \bf p^j})$ with respect to that parameter to $0$ and solving for $p^j(i)$.\\
This is an iterative process which can be started with an arbitrary 
choice for the parameters in the first step, ${\bf p^o}$. The EM algorithm 
assure us that at every step the likelihood $\mathcal{L}\rm({\bf p})$ is increased. 
The method will converge to the optimal set of parameters, ${\bf p}$, in a number of 
steps which will depend on the nature of the problem and on the initial election for the first 
iteration, ${\bf p^o}$.\\

In this paper, we will consider a simple case where we allow the covariance in Fourier space   
of the CMB signal (or power spectrum) to be a free parameter. \\
We will apply the EM algorithm in order to determine that free parameter 
(the power spectrum of the signal). 
This case can be extended to include more free parameters in the analysis as it is shown 
in Delabrouille et al. (2002). \\

\section{EM applied to CMB experiments}\label{em_cmb}
If we know the frequency dependence of one of the components, then we can express 
the signal at a given frequency as: 
\begin{equation}
{\bf S}({\bf n},\nu) = {\bf A}({\bf n},\nu)\otimes \Big( {\bf f}(\nu) s({\bf n}) \Big)
\end{equation}
where $s({\bf n})$ is the spatial pattern of the signal we want to estimate, 
${\bf f}(\nu)$ is the frequency dependence of the signal (including the band-with), and 
${\bf A}({\bf n},\nu)\otimes$ accounts for the convolution with the antenna beam of the experiment 
and its frequency response. \\
The data at different frequencies, ${\bf d}(\nu)$, can be expressed as a sum 
of the signal plus some other contributions. 
\begin{equation} 
{\bf d}({\bf n},\nu) = {\bf S}({\bf n},\nu) + {\bf \xi}({\bf n},\nu).
\label{eq_hypo_RealSpace}
\end{equation} 
The residual ${\bf \xi}({\bf n},\nu)$ includes all the other components, 
i.e. Galactic and extragalactic components convolved with the corresponding
beam plus the corresponding instrumental noise. 
Due to the antenna and frequency convolution in ${\bf A}({\bf n},\nu)\otimes$, 
it is easier to work in Fourier space where this convolution is just a product 
of the Fourier modes which are uncorrelated provided the field is  homogeneous 
and isotropic. 
Therefore, the previous equation should be rather expressed in Fourier space as: 
\begin{equation} 
d_{\nu}({\bf k}) = A_{\nu}( k ) f_{\nu} s({\bf k}) + \xi_{\nu}({\bf k}).
\label{eq_hypo}
\end{equation} 
If the experiment has $m$ different channels, then we have an observation 
for each channel. That is, we have the vector of observations 
${\bf d} = (d_1,d_2,...,d_m)$ for each Fourier mode ${\bf k}$. And we can write:
\begin{equation}
{\bf d} = {\bf R} s + {\bf \xi}
\label{eqn_vec_d}
\end{equation}
where, ${\bf R}$ is the response vector containing $m$ elements.
Each element of ${\bf R}$ is just the product of the antenna and the frequency
response of the instrument times the frequency dependence of the signal,
\begin{equation}
R_{\nu} = A_{\nu} f_{\nu}. 
\end{equation}
In the case in which the signal is the CMB then $f_{\nu} = 1 \ \ \forall ~\nu$. 
From eqn. \ref{eqn_vec_d} we can express the residual as
\begin{equation}
{\bf \xi} = {\bf d} - {\bf R}s. 
\label{eqn_residual}
\end{equation}

The basic starting point when applying EM is to define the pdf of the complete data. 
In our case the complete data are all the components plus the multifrequency 
observations $\mathbf d$. 
We will focus on the problem in which the complete data can be divided in 
just two elements.
The two elements are the CMB signal we want to estimate $s$ and the 
observed data $\mathbf d$ (or equivatently the residual $\xi$, noise plus rest 
of components; see below). 
By doing this, 
the nature of our problem reduces drastically since we only have to make assumptions about two elements. \\
Furthermore, the residual and the signal can be considered as independent. 
In terms of the probability, the pdf of the complete data is:
\begin{equation}
{\bf f}({\bf x}|{\bf p}) = {\bf P}(s,{\bf d}|{\bf p}) = {\bf P}(s,{\bf \xi}|{\bf p} ) = {\bf P}(s|{\bf p_A})
{\bf P}({\bf \xi}| {\bf p_B})
\label{eq_complete_pdf}
\end{equation}
where ${\bf p} = \bigcup {\bf p_A} {\bf p_B}$ with ${\bf p_A}$ and ${ \bf p_B}$ unknown 
parameter sets for the pdf's of $s$ and ${\bf \xi} $. The second equality follows from
the transformation $(s,{\bf d}) \rightarrow (s,{\bf \xi})$ whose Jacobian is 
equal to one (see equation \ref{eq_hypo}). 
It is important to note that, in oppossition to other methods, our assumption 
about independence of the elements ($s$ and ${\bf \xi}$) is
well justified. No relation is expected between the CMB and the other foregrounds (except maybe the 
SZ effect which could be weakly related to the CMB through the ISW effect). \\

Once we have established that the two elements are independent, 
we need a clue about the specific form of the individual 
probabilities, ${\bf P}(s|{\bf p_A})$ and ${\bf P}({\bf \xi}|{\bf p_B})$. 
If the CMB is close to a Gaussian variable (and we know it is), 
then the pdf of the CMB in Fourier space can be approached by:
\begin{equation}
{\bf P}(s|{\bf p_A}) \equiv {\bf P}(s) \propto P(k)^{-1} exp(- \frac{s^2}{P(k)})
\label{eq_Ps}
\end{equation}
where $P(k)$ is the power spectrum of the CMB: the parameter to be estimated. \\
The pdf of the residual (${\bf \xi}$) can be modelled as 
\begin{equation}
{\bf P}({\bf \xi}|{\bf p_b}) \equiv {\bf P}({\bf \xi})
\propto exp(-\boldmath{\xi} C^{-1}\boldmath{\xi}^{\dag})
\label{eq_likelihood}
\end{equation}
where $C^{-1}$ is the inverse of the correlation matrix of the residual. $C$ is 
an $n\times n$ matrix ($n =$ number of channels). 
This correlation matrix is not diagonal if there are some correlations in the residuals 
between the different frequency channels. The elements of $C$ are 
the power (diagonal) and cross-power (off-diagonal) spectra of
the residuals at each channel which can be given as a function of the $P(k)$ (see below).\\
We have assumed in equation (\ref{eq_likelihood}) that the residuals are 
Gaussian distributed. This assumption is not as strong as assuming that each 
component is Gaussian distributed. The sum of several non-Gaussian 
distributions tends to a more Gaussian one. In the limit of a sum of infinite
independent components the central limit theorem assures that the resulting 
distribution is Gaussian. In any case Gaussianity can be a good approach at 
scales where the instrumental noise dominates the residuals. The most critical
situations for the validity of this assumption appear at large and small 
scales since, at least for some channels, they are dominated by the Galactic 
components and the compact sources, 
respectively. Solutions for these problems will be proposed in the next 
sections.  

In the previous relations for the individual pdf's appear the power 
spectra of the signal and the residual. In a real situation  
these quantities are unknown. This is the reason why we apply EM.  
We can introduce the unknown CMB power spectrum, $P(k)$, as the uncertainty in our problem.
In order to apply EM and estimate this free parameter we have to calculate the 
expected value $Q({\bf p} |{\bf p^j})$ which is just an integral:
\begin{equation}
Q({\bf p | p^{j}}) = \int log {\bf P}(s,{\bf d}|{\bf p}) 
{\bf P}(s|{\bf d},{\bf p^j}) ds.
\label{eq_Q}
\end{equation}
The only thing we need to know to calculate $Q({\bf p} |{\bf p^j})$ are the probabilities 
${\bf P}(s|{\bf d})$ and ${\bf P}(s,{\bf d})$. 
By Bayes we know that:
\begin{equation}
{\bf P}(s|{\bf d}) = \frac{{\bf P}(s,{\bf d})}{{\bf P}({\bf d})}
\end{equation}
where ${\bf P}({\bf d})$ can be obtained just by marginalising ${\bf P}(s,{\bf d})$ over $s$: 
\begin{equation}
{\bf P}({\bf d}) = \int {\bf P}(s,{\bf d}) ds
\end{equation}
and ${\bf P}(s,{\bf d})$ can be obtained through the chain rule:
\begin{equation}
{\bf P}(s,{\bf d}) = {\bf P}(s,{\bf \xi}) J\left ( \frac{s,{\bf \xi}}{s,{\bf d}} \right ).
\end{equation}
By taking equation (\ref{eq_hypo}), one can
prove that the Jacobian in the previous 
expression is equal to one, as it was commented above, so we have:
\begin{equation}
{\bf P}(s,{\bf d}) = {\bf P}(s,{\bf \xi}) 
\end{equation}
where (see equation \ref{eq_complete_pdf}), 
\begin{equation}
{\bf P}(s,{\bf \xi}) \propto P(k)^{-1} exp(- \frac{s^2}{P(k)})exp(-{\bf \xi} C^{-1}{\bf \xi}^{\dag}) 
\end{equation}
where the term, $P(k)$, accounts for our free parameter in ${\bf P}(s|{\bf p_A})$
(see equation \ref{eq_Ps}) and the second term was taken from equation (\ref{eq_likelihood}).
We remark that the covariance matrix $C$ also depends on the free parameter $P(k)$, as it will be
discussed below.

Now it is easy to compute the terms, ${\bf P}({\bf d})$, ${\bf P}(s|{\bf d})$, and finally 
$Q(P(k)|P(k)^j)$:
\begin{equation}
{\bf P}({\bf d}) \propto \frac{exp\Big( - {\bf d}C^{-1}{\bf d}^{\dag}  + 
              ({\bf d}C^{-1}{\bf R}^{\dag})^2 /F_1 \Big)}
               {P(k)|C|^{1/2}F_1}
\end{equation}
where we have expressed ${\bf \xi}$ in terms of ${\bf d}$ and $s$ (see eqn. \ref{eqn_residual}). 
\begin{equation}
{\bf P}(s|{\bf d}) \propto 
F_1 exp \left ( - F_1 \Big| s - \frac{{\bf d}C^{-1}{\bf R}^{\dag}}{F_1}\Big|^2  \right )
\label{eq_prob_s_cond_d}
\end{equation}
where, 
\begin{equation}
F_1 = {\bf R}C^{-1}{\bf R}^{\dag} + P(k)^{-1}.
\end{equation}
Finally the expected value of the log of the complete pdf (equation 
\ref{eq_Q}) is:
\begin{eqnarray}
Q(P(k)|P(k)^j) &=& -log(P(k)) - \frac{1}{2}log|C| - {\bf d}C^{-1}{\bf d}^{\dag} \\ \nonumber
         &-& F_1<|s|^2> + 2\mathbf{d}C^{-1}{\bf R}^{\dag}<s> + const. 
\label{eq_Q_P}
\end{eqnarray}
where :
\begin{equation}
<s> = \frac{{\bf d}C^{-1}{\bf R}^{\dag}}{F_1}
\label{eq_search_engine_EM}
\end{equation}
\begin{equation}
<|s|^2> = \frac{1}{F_1}\left ( 1 + \frac{|{\bf d}C^{-1}{\bf R}^{\dag}|^2}{F_1}\right ).
\label{eq_variance_EM}
\end{equation}
Equation (\ref{eq_search_engine_EM}) is nothing more than multifrequency 
Wiener 
filter computed with the CMB power spectrum at iteration $j$. This is an 
expected result since Gaussianity for the signal and 
residuals have been assumed. Similarly, the signal variance at each $k$ 
$<|s|^2>$ is also computed with $P(k)^j$ at iteration $j$. At this point it is 
important to note that in eq. (\ref{eq_Q_P}) the 
dependence on $P(k)$ appears explicitly in the first term and also in the 
fourth 
one through the function $F_1$. Moreover, since we do not know the residuals 
the $P(k)$ dependence is also present in the other terms through the residual 
covariance matrix, 
\begin{equation}
C^{\nu\mu} = P_{\bf \xi}^{\nu\mu} (k) =  P_{\bf d}^{\nu\mu}(k) - R^\nu 
R^\mu P(k) 
\end{equation}
where  $P_{\bf \xi}^{\nu\mu}(k)$ and $ P_{\bf d}^{\nu\mu}(k)$ are the residual
and data cross-power spectra after convolution with the corresponding 
antenna response whereas $P(k)$ is the unconvolved CMB power spectrum to be 
determined.

In the maximisation step, the value of $P(k)^{j+1}$ that maximises $Q(P(k)|P(k)^j)$ 
should be found. Finding the maximum is not a trivial task given the 
complicated dependence on $P(k)$. In order to facilitate the finding we 
consider 
the  residual covariance matrix, $C^{\nu\mu}$, constant in the M-step (but 
changing with $P(k)$ in the E-step). 
With this simplification, 
we can easily find the analitical solution for the maximum by differentiating 
$Q(P(k)|P(k)^j)$ with respect to $P(k)$ and equating to zero. We find the not 
so surprising result: 
\begin{equation}
P(k)^{j+1} = <|s|^2>.
\label{eq_power_EM}
\end{equation}
The previous equation is the main result of EM. $<|s|^2>$ is given by equation
(\ref{eq_variance_EM}) computed with the value of $P(k)$ at iteration $j$.
By iterating equation (\ref{eq_power_EM}), which only depends on the data, 
the response vector and the CMB power spectrum obtained in the previous 
iteration, we can get an estimate of the power spectrum of the underlying 
signal. The final step after the
EM estimation of the signal power spectrum at each $k$, is to
recover the CMB map by applying eqn. \ref{eq_search_engine_EM}, the 
multifrequency Wiener filter derived within the EM framework.

Finally, this method can be easily extended to determine jointly the CMB and the SZ
emissions, since they are almost uncorrelated and the frequency dependence 
of the SZ is also very well known. In addition, the method can be implemented with
more realistic (non-Gaussian) pdfs to model the signal distributions as well as the
residuals. We will study these extensions of the method in a future work.

\section{Results}

\begin{figure*}
   \begin{center}
	\includegraphics[scale = 0.9]{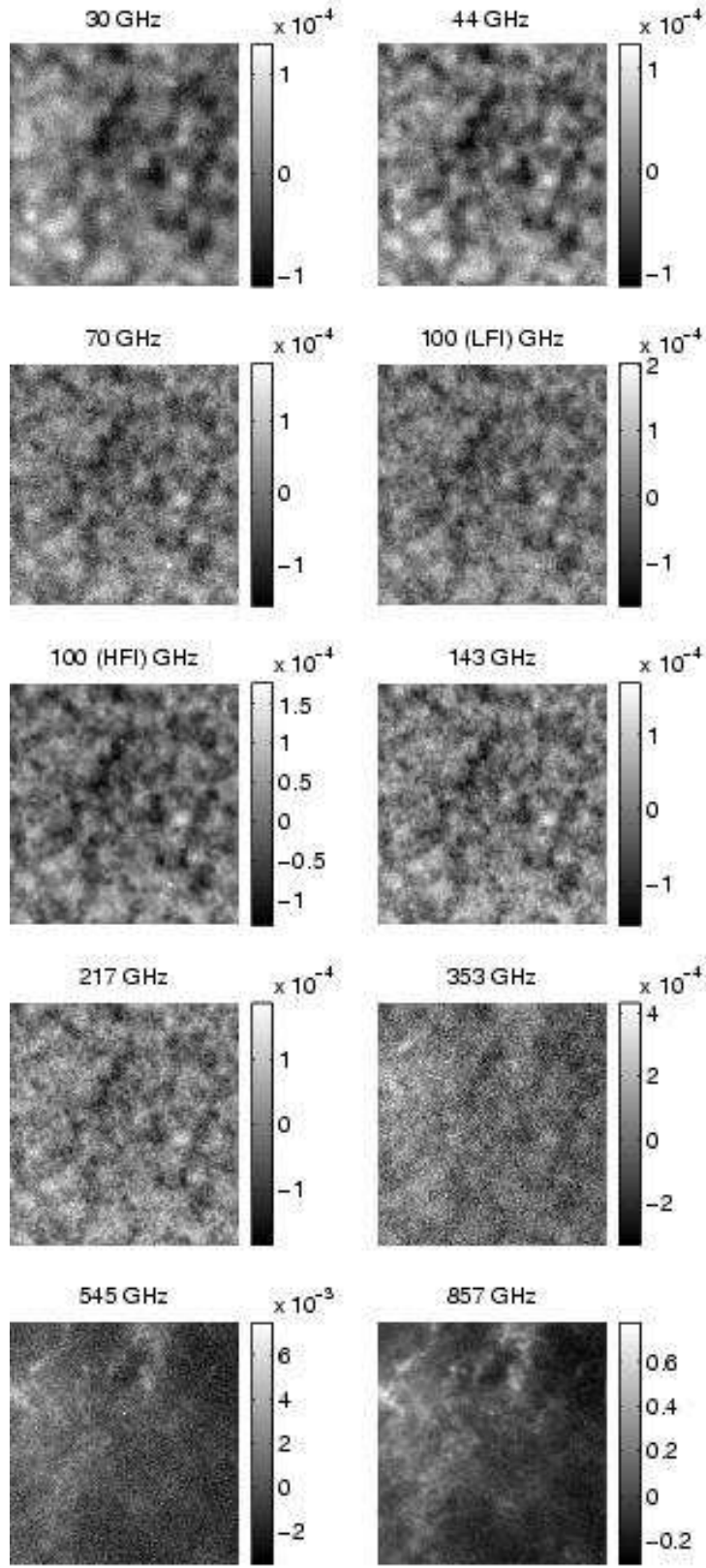}
   \caption{
Realistic simulations of the data, covering a $12.8^o\times 12.8^o$ patch, 
expected from the 10 Planck channels. The RMS amplitude (at 353 GHz
	and in $\Delta T/T$ thermodinamic temperature units)
	is $4.26\times10^{-5}$ for the CMB, 
	$5.16\times10^{-6}$ for the PS emission, $3.93\times10^{-6}$
	for the SZ contribution, $5.82\times10^{-5}$ for the thermal
	dust, $3.23\times10^{-7}$ for the free-free radiation, 
	$1.46\times10^{-6}$ for the synchrotron emission and it is
	almost negligible in the case of the spinning dust (see text).
           }
   \label{fig_maps}
   \end{center}
\end{figure*}

In this section, we present the results that have been obtained for the
CMB power spectrum determination. Firstly, we briefly describe the
simulated data that have been used in this work, accounting for
the most important contributions to the microwave sky.
Then, we present our CMB power spectrum estimation.
An improvement at small
scales is achived by using a very useful tool
for subtracting the brightest point sources: the Mexican Hat Wavelet.

Finally, we applied our estimator to an ideal data set,
where the Gaussian assumption for the residuals is satisfied.
This exercise is highly interesting,
since it shows the robustness of our estimator: the results
achieved in this idealistic case are only slightly better than the ones
obtained in the reslistic situation.

\subsection{Simulated data}

\begin{table}
   \begin{center}
         \begin{tabular}{|c|c|c|c|}
	 \hline
	 Frequency & FWHM & Pixel size & $\sigma_{noise}$ \\
	 (GHz) & (arcmin) & (arcmin) & $(10^{-6})$ \\
	 \hline
	 857 & 5.0 & 1.5 & 22211.10 \\
	 \hline
	 545 & 5.0 & 1.5 & 489.51 \\
	 \hline
	 353 & 5.0 & 1.5 & 47.95 \\
	 \hline
	 217 & 5.5 & 1.5 & 15.78 \\
	 \hline
	 143 & 8.0 & 1.5 & 10.66 \\
	 \hline
	 100 (HFI) & 10.7 & 3.0 & 6.07 \\
	 \hline
	 100 (LFI) & 10.0 & 3.0 & 14.32 \\
	 \hline
	 70 & 14.0 & 3.0 & 16.81 \\
	 \hline
	 44 & 23.0 & 6.0 & 6.79 \\
	 \hline
	 30 & 33.0 & 6.0 & 8.80 \\
	 \hline
      \end{tabular}
      \caption{\label{table:instrument}
Experimental constrains at the 10 Planck channels. The
      antenna FWHM is given in column 2 for the different frequencies
      (a Gaussian pattern is assumed in the HFI and LFI
      channels). Characteristic pixel sizes are shown in column 3. The
      fourth column contains information about the instrumental noise
      level, in $\Delta T/T$ per pixel.}
    \end{center}
\end{table}

To show the power of our approach we will apply our estimator 
(eqn. \ref{eq_power_EM}) to 
realistic simulations. They consider the expected levels of the instrumental
white noise and resolution characteristics of the 
10 Planck mission channels at 30, 44, 70, 100 (Low and High Frequency
Instrument, LFI and HFI),
143, 217, 353, 545 ad 857 GHz (see Table~\ref{table:instrument}).
In addition, we use state of the art simulations 
of the different galactic and extragalactic components, together with
the pure CMB signal. 

Within the first ones, we have taken into account the 
synchrotron, free-free, spinning and 
thermal dust contributions. The first one has been simulated using the all-sky
pattern provided by Giardino et al. (2002) including both,
temperature and spectral index templates. The free-free emission is
very poorly known. Recent experiments focusing on the H-$\alpha$
detection, like Southern H-$\alpha$ Sky Survey
(SHASSA, Reynolds \& Haffner, 2000) and the Wisconsin H-$\alpha$ Mapper
project (WHAM, Gaustad et al. 2001), will produce all-sky surveys that
could be used as free-free templates. However, since these data are not
available at the present time, we have used the correlation between
dust and free-free emissions proposed by Bouchet et al. (1996) as
spatial template, and a power law, $I_{\nu} \propto {\nu}^{-0.16}$, to
describe the frequency dependence.
We have also included the spinning dust emission --proposed by Draine \&
Lazarian (1998) as a possible explanation for the anomalous Galactic
emission found by CMB experiments like COBE (Kogut 1999) or Saskatoon
(Oliveira-Costa et al. 1997). Very recently, a tentative confirmation of
that emission has been claimed (Finkbeiner et al. 2002).
The simulation of that emission has been carried out using the
frequency dependence proposed by 
Draine \& Lazarian (1998)\footnote{Data provided by the authors.}
and using the thermal dust component as an spatial
template, since both dust emissions are strongly correlated through
the neutral hydrogen column density ($N_H$).
Finally, thermal dust emission has been also simulated, using the best
model proposed by Finkbeiner et al. (1999) to fit the FIRAS, IRAS and DIRBE
data, consisting of two grey-bodies with mean temperatures of 16.2 K and
9.4 K and emissivities 2.70 and 1.67, respectively.
The simulated extragalactic foregrounds are the thermal Sunyaev-Zel'dovich
effect (SZ) and radio and infrared point
sources.
The SZ was performed following the Diego et al. (2001) model for a
flat $\Lambda$CDM Universe with $\Omega_m = 0.3$ and
$\Omega_{\Lambda}$ = $0.7$. 
The point sources correspond to the Toffolatti et al. (1998) model
for the same Universe, including radio flat-spectrum and infrared
sources. We refer to that paper for details.
Finally, the CMB signal was simulated using the $C_l$'s provided by the
CMBFAST code (Seljak \& Zaldarriaga, 1996) for the same cosmological
model and assuming Gaussian fluctuations. We would like to remark that
synchrotron, thermal dust and the point source emissions have been
simulated using a frequency dependence which varies with the sky position.
The simulated maps covering a patch of the sky of $12.8^o\times 12.8^o$, as 
seen by the 10 Planck channels, are shown in figure 
\ref{fig_maps}.

\begin{figure}
   \begin{center}
	\includegraphics[angle = 0,width=8cm]{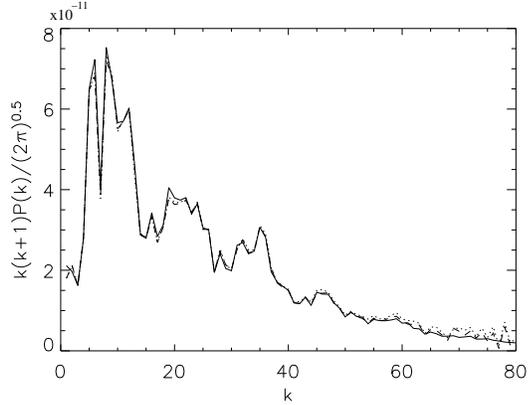}
   \caption{
            Recovered power spectrum by our estimator. The solid line is the 
	    true power spectrum and the dotted and dashed lines represent the 
	    recovered power spectrum obtained by our estimator directly from 
	    the data and the one recovered after previous subtraction of the 
	    brightest point sources, respectively. P(k) is in $\Delta T/T$ units. Notice that the multipole $l$ is related to the wave number $k$ through 
$l\approx 28 k$.
           }
   \label{fig_recover_power}
   \end{center}
\end{figure}

\begin{figure}
   \begin{center}
	\includegraphics[angle= 0,width=8cm]{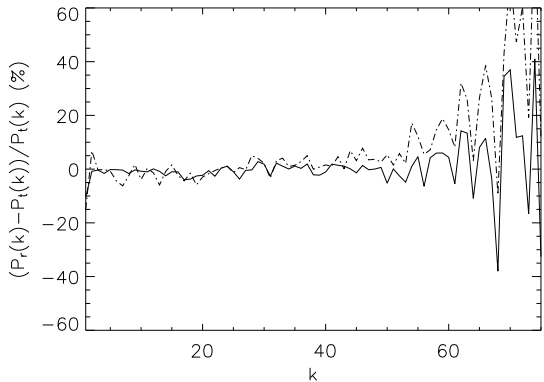}
   \caption{
            Relative error in \% of the recovered power spectrum by our 
	    estimator (dash-dotted line). For comparison we have plotted the 
	    errors obtained by our estimator in the ideal situation when only 
	    CMB and instrumental white noise are present in the data (see 
	    text).
           }
   \label{fig_RelatErr}
   \end{center}
\end{figure}
 
%
%
\begin{figure}
   \begin{center}
	\includegraphics[angle = 0,width=8cm]{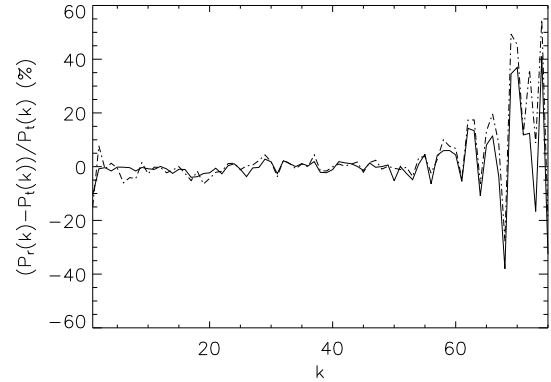}
   \caption{
            Relative error in \% of the recovered power spectrum by our 
	    estimator after the brightest point sources have been removed in 
	    all the channels (dash-dotted line). For comparison in the solid 
	    line we have plotted
	    the errors obtained by our estimator in the ideal situation when 
	    only CMB and instrumental white noise are present in the data (see 
	    text).
           }
   \label{fig_RelatErr_NoPS}
   \end{center}
\end{figure}

\subsection{A simple choice for the correlation matrix}

In order to estimate the power spectrum of the CMB through 
eqn. (\ref{eq_power_EM}) we still have to iterate a relatively complex 
expression containing a cocient of polynomials of order twice the 
number of channels and involving many terms.
To further simplify the convergence of that equation we will only 
consider the diagonal terms of the covariance matrix, making zero the other 
non-diagonal elements. This assumption does not take into account
the correlations among the residuals at different channels due to the 
Galactic components and compact sources.
The former are reflected at the low $k$-modes and their effect can be
reduced by performing a simple Galactic subtraction as the one proposed by
Diego et al. (2002). The later 
have their impact at intermediate (e.g. radio sources
observed in the lowest resolution channels)
and large $k$-modes (e.g. infrared sources). 
However, as it will be shown below, once the brightest point sources
are subtracted from the maps, the contribution of the remanent point sources
is low enough to make their effect significantly smaller.

In figure \ref{fig_recover_power} we present our results for the 10 Planck 
channels using the 
data obtained from the combination of the simulated components as described 
above (figure \ref{fig_maps}).
The comparison of the recovered spectrum with the 
true one
shows a good agreement until $k$ values larger than the one corresponding to 
the best antenna channels and the results start to be dominated by the 
deconvolved noise.    

In order to better appreciate the differences, we show in figure 
\ref{fig_RelatErr} the relative error,
$(P_{r}(k) - P_{t}(k))/P_{t}(k)$ \%, where $P_{r}(k)$ is the
estimator of equation \ref{eq_power_EM}.
  
In both large and relatively small scale range (small and
relatively large $k$) our 
estimator finds a reasonably good 
estimation of the true power spectrum. The error is small (less than 10 \%) 
up to scales $k \approx 55$ (which corresponds to $l\approx 1500$). The 
effect of only considering the diagonal terms of the residual covariance 
matrix is small (see also the comparison with the ideal case below). This fact 
proves the applicability of our assumption for the microwave data 
since our method is not very sensitive to the cross-correlations of the 
different channels. 

At very small scales ($k > 45$ or $l > 1300$) the signal is below the noise 
level and the error bars grow quickly. Our estimator shows a clear bias toward
higher values (we recover more power than the true one). This implies that 
some of our assumptions are wrong at these small scales.
The Gaussianity assumption is a good approximation in Fourier space at large
$k$-modes (because of the dilution in many modes
of the clear non-Gaussian features appearing in the real space).
However, the uncorrelation assumption among frequency channels is
a source of bias at these scales, since the bright point sources
are strongly correlated.
We have checked 
this point by removing the brightest point sources (but leaving the galaxy clusters and the weak 
point sources in the residual). We have used the Mexican Hat Wavelet
(MHW) technique proposed by Vielva et al. (2001a)
for removing point sources above the fluxes determined by the so-called $50\%$ error
criterion used in that paper. These fluxes correspond to those
detection limits above which the maximum percentage of spurious
detection is $5\%$, being a detection spurious if the error in the
amplitude estimation is larger than $50\%$.
After the subtraction, we
applied our estimator to these new {\it point-source-free} maps.
The final result is shown in 
figure \ref{fig_RelatErr_NoPS}.
The effect of removing the brightest point sources is evident now. 
This result clearly shows how the bias can be corrected by removing
the compact sources. 
The relative error does not change significantly after removing the
point sources but the estimator 
is able to go up to scales $k = 60$ ($l \approx 1700$) with relative error 
$< 10 \%$. 
The error is smaller than $50 \%$ below $k = 75$ ($l \approx 2100$). 

\begin{figure*}
   \begin{center}
	\includegraphics[angle = 270, width=16cm]{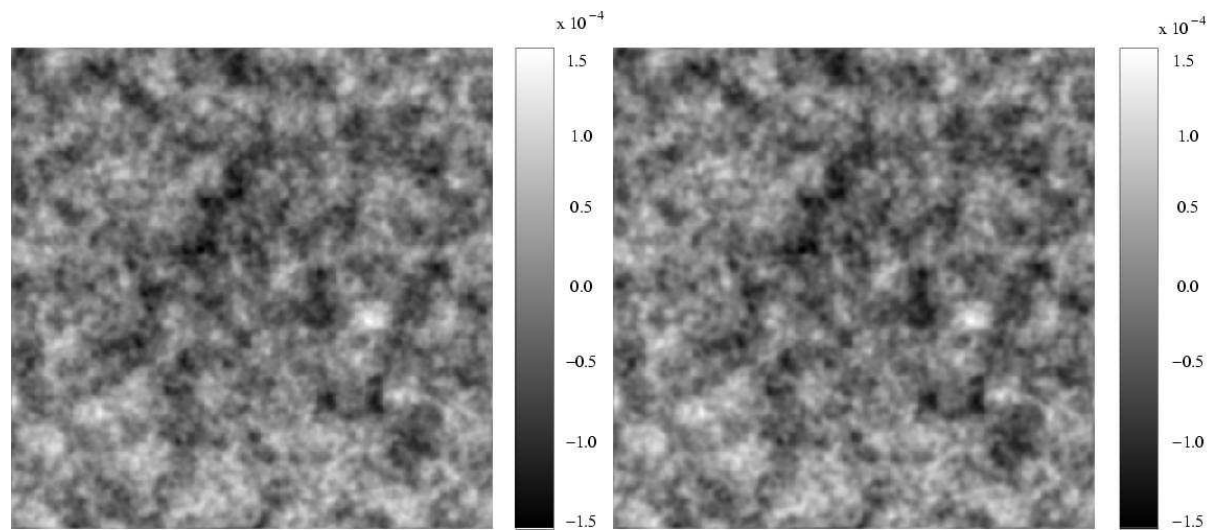}
   \caption{
   Left: CMB map reconstruction by our method
   (equation \ref{eq_search_engine_EM}) 
   using the recovered power
   spectrum obtained after previous subtraction of the brightest point sources.
   Right: Input CMB map. Both maps have been convolved with a Gaussian beam
   of FWHM = 5 '
   }
   \label{fig_map_reconstruction}
   \end{center}
\end{figure*}

\subsection{Comparing with an ideal situation}\label{section_idealcase}

It is interesting to answer whether or not the above results are close to the most optimistic 
case where all our assumptions are right (Gaussianity and $C$ diagonal). 
We will check this point in the simple (but unrealistic) case where our data consist only of 
CMB signal plus the corresponding instrumental noise for each channel. 
In this simple case, our assumption of Gaussianity is right since both CMB and noise 
are Gaussian. The correlation matrix of the residual, $C$, will be exactly
diagonal with the elements being the power spectrum of the data minus the CMB
power spectrum convolved with the antenna corresponding to each channel.  

In figures \ref{fig_RelatErr} and \ref{fig_RelatErr_NoPS} we show the 
performance of the EM estimator in this 
ideal case (CMB plus noise) after combining the 10 Planck channels.  
The result is surprisingly very similar to the one obtained in the realistic 
case where the brightest point sources have been subtracted (figure 
\ref{fig_RelatErr_NoPS}). 
This test shows how the EM estimator is extremely 
robust and can give results very close to the most optimal one. 

\subsection{Map reconstruction}
\begin{figure}
   \begin{center}
	\includegraphics[angle = 0,width=8cm]{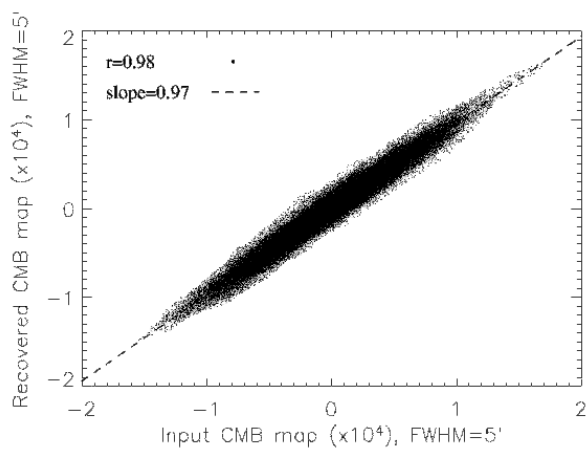}
   \caption{
            Correlation between the input and recovered CMB maps. The best strightline fit
	    is also plotted.
   }
   \label{fig_correl_coefficient}
   \end{center}
\end{figure}

Using the expression given in equation~\ref{eq_search_engine_EM}
and the estimated power spectrum $P_r (k)$, we are able to recover the
CMB map. As it was pointed out in Section~\ref{em_cmb}, due to the Gaussianity
assumptions, equation~\ref{eq_search_engine_EM} is nothing more than multifrequency
Wiener filter (MWF). Hence, our map reconstruction is a MWF focused on
the CMB recovery. Let us remark that this approach differs from the traditional
MWF already used in the microwave sky recovery (e.g., Tegmark \& Efstathiou 1996,
Bouchet \& Gispert 1999 and Hobson et al 1998). These works were focused on the
all components separation, whereas our approach is devoted to reconstruct just the
CMB signal. Even more, whereas the traditional approaches require previous knowledge of the
power spectra (not only the CMB one, but also the power spectra of the foregrounds),
the CMB  power spectrum is a result of the blind EM method.
This and the non requirement of any prior knowledge
about the frequency dependence of the components, are the strongest points of
the blind methods. In figure~\ref{fig_map_reconstruction} we present the
CMB map reconstruction (left) together with the input one (right). Both maps have been
convolved with the best Planck resolution (FWHM = 5').
The correlation between the input and recovered CMB maps is very good 
$r = 0.98$
(figure~\ref{fig_correl_coefficient}), whereas the slope of the best 
straightline fit is 0.97.
The residual CMB is dominated by the Galactic dust emission, since it is the most important foreground
at low $k$-modes. Hence, by performing a simple
dust subtraction as the one proposed by Diego et al. (2002)
--which basically consists in using the 857 GHz channels as a dust
template to be subtracted from the others channels--
a significant improvement in the
CMB map reconstruction is achieved, being the RMS of the residual map lower 
than $15\%$.

\section{Conclusion}\label{section_conclusion}
We have presented an efficient estimator (eqn. \ref{eq_power_EM}) of the CMB 
power spectrum. 
Our estimator is based on the EM algorithm
and does not make use of any prior information
whatsoever about any of the components. This
renders satisfactory results, since we are able to recover the power spectrum 
up to $l\approx 1500$ 
with less than 10\% error. This limit can be improved if we remove the 
brightest compact 
sources, for instance applying the MHW technique poposed by Vielva et
al. (2001a). After point source removal the estimator can recover 
the power spectrum up to $l\approx 1700$ with less than 10\% error and 
up to $l\approx 2100$ with less than 50\% error.

Our results are very close to the optimal case when all our assumptions are
satisfied (only CMB and 
instrumental noise are included in the simulations), as can be seen in 
figure~\ref{fig_RelatErr_NoPS}. 
There are several CMB power spectrum estimations in the literature,
given for different component separation techniques, like Bayesian
methods (MEM: Hobson et al. 1998 and Stolyarov et al. 2002; and MWF:
Tegmark \& Efstathiou 1996 and Bouchet \& Gispert 1999)
and blind source separation algorithms (ICA and FastICA:
Baccigalupi et al. 2000 and Maino et al. 2002; and Multi-Detector
Multi-Component analysis: Delabrouille et al. 2002).
This work provides a CMB power spectrum
that is comparable to or even better than the previous ones, being more
robust than those, since previous knowledge about the
underlying components is not requiered (contrary to the Bayesian methods)
and simulations 
used in previous works are significantly more idealised than the ones
used in this work (e.g., lack of spatial variation of the
frequency dependence for several components;
absence of some of the components, like point
sources or some Galactic foregrounds).
On the other hand, our simulations account for the previous limitations including
all the major microwave emissions and allowing for spatial variation of the
frequency dependence. 
Finally, our method is able to reconstruct the CMB map using the estimated power spectrum
and the MWF given in equation~\ref{eq_search_engine_EM}. Let us remark that the MWF is
obtained when the Gaussianity hypothesis is assumed, whereas the EM algorithm provides
a more general procedure for component separation. 
That hypothesis
will be explored in a future work by including more
realistic distributions for the different components. In addition, we are working on a jointly
determination of the cosmological signals (CMB, SZ and point sources)
using the EM algorithm and a point source detection tool. Finally, an 
extension of this algorithm to the sphere is straightforward and will be 
studied soon.

\section*{Acknowledgements}
We thank J. Delabrouille for a critical reading of a previous version of the
manuscript which resulted in a substantial improvement.
We thank L. Toffolatti for providing us with the point source simulation, 
and B. T. Draine and A. Lazarian for providing
us with the emissivity predicted by their spinning dust model.
We appreciate interesting discussions with J. L. Sanz, R. B. Barreiro,
D. Herranz and V. Stolyarov.
We thank the RTN of the EU project HPRN-CT-2000-00124 for partial finantial 
support.
EMG and PV acknowledge partial financial support from the Spanish MCYT
projects ESP2001-4542-PE and ESP2002-04141-C03-01.
This research has been supported by a Marie Curie Fellowship 
of the European Community programme {\it Improving the Human Research 
Potential and Socio-Economic knowledge} under 
contract number HPMF-CT-2000-00967.
PV acknowledges support from a fellowship of Universidad de Cantabria.



\bsp
\label{lastpage}
\end{document}